\documentclass{desyproc}

\begin{document}
\title{Detecting Sub-eV Scale Physics by Interferometry}

\author{{H. Tam and Q. Yang}\\[1ex]
Department of Physics, University of Florida, Gainesville, FL 32611}

\contribID{familyname\_firstname}

\desyproc{DESY-PROC-2012-04}
\acronym{Patras 2012} 
\doi  

\maketitle

\begin{abstract}
We propose a new interferometry-based experiment to detect sub-eV scale particles such as axion-like particles (ALPs).
\end{abstract}

ALPs are predicted to exist in string theory. Pseudoscalar ALPs couple to photons as axions do and scalar ALPs couple to photons via a $aF_{\mu\nu}F^{\mu\nu}$ term in the Lagrangian. In general, there is not a priori relationship between the mass and the couplings of ALPs, so their parameter space is a lot less constrained compared to axions. In addition, many ultralight hypothetical particles can also couple to photons as ALPs do, such as sub-eV hidden Higgs. These particles are predicted naturally from the hidden sector of theories embedding the Standard Model. Unlike for the TeV scale physics, particle colliders may be not the best ways to detect these weakly interacting sub-eV particles (WISPs) so a large number of non-collider experiments worldwide are currently actively searching for WISPs e.g. BFRT, BMV, ADMX, CAST, PVLAS, GammeV, CARRACK, ALPS(at DESY), OSQAR(at CERN), etc. As of today these particles remain undetected. In this paper, we propose a new experimental method~\cite{tamyang} based on interferometry to detect the WISPs.

photons and ALPs are mixing in an external magnetic field.  The probability of conversion from photons to ALPs $\eta$ can be obtained from the cross section of this process, which was first done in~\cite{Sikivie:1983ip} and is given by
\begin{equation}\label{conversion1}
\eta_{\gamma \rightarrow a} = \frac{1}{4 v_a}(g_{a\gamma\gamma} B L)^2\left(\frac{2}{qL}\sin{\left(\frac{qL}{2}\right)}\right)^2,
\end{equation}
where $v_a$ is the velocity of the ALP, $B$ is the magnetic field, $L$ is the length of the conversion region, and $q$ is the momentum transferred to the magnet. Since $m_a \ll \omega_{\gamma}\sim {\rm eV}$, the frequency of the laser beam photons, $v_a \sim 1$ and $q=m_a^2/2\omega_{\gamma}$. For $L \sim 10 $m, $m_a\sim 10^{-6}{\rm eV}$, this also implies that $qL \sim 10^{-5} \ll 1$. So \eqref{conversion1} can be approximated by
\begin{equation} \label{conversion2}
\eta_{\gamma \rightarrow a} \approx \frac{1}{4}(g_{a\gamma\gamma}B L)^2.
\end{equation}
If we use $B \sim 10 $T, $L \sim 10 $m, and $g_{a\gamma\gamma} \sim 10^{-15}$ GeV$^{-1}$, the probability of photon-axion conversion is of $\mathcal{O}(10^{-26})$.

After the conversion, the amplitude $A$ of the photon is reduced to $A-\delta A$, where
\begin{equation}\label{daaxion}
\delta A_{\gamma \rightarrow a} = \frac{A \eta_{\gamma \rightarrow a}}{2} \approx \frac{g_{a\gamma\gamma}^2 B^2 L^2 A}{8}.
\end{equation}
Equation \eqref{daaxion} is valid when $m_a\ll m_0\equiv\sqrt{2\pi\omega_{\gamma}/L}$, which is about $10^{-4}{\rm eV}$ for given $L$ and $\omega_{\gamma}$. If $m_a$ is larger than $m_0$ the power loss effect decreases rapidly. When $m_a\gg m_0$, $\delta A_{\gamma \rightarrow a} \sim {g_{a\gamma\gamma}^2 B^2 L^2 A}({m_0/ m_a})^4$.

Under weak mixing assumption, the additional phase acquired $\delta \theta$ is approximately\cite{Raffelt:1987im}
\begin{equation}\label{dthetaaxion}
\delta \theta\approx \frac{g_{a\gamma\gamma}^2 B^2 \omega_{\gamma}^2}{m_a^4}({m_a^2L\over 2\omega_{\gamma}}-{\rm sin}({m_a^2L\over 2\omega_{\gamma}})).
\end{equation}
The effect of the phase shift is negligible in comparison with $\delta A/A$ when $m_a\sim 10^{-6}{\rm eV}$. When $m_a\gg m_0$ the effect of the phase shift is comparable or even bigger than $\delta A/A$. However, as we will show, the signal due to $\delta A/A$ registered by the detector is of first order and the signal due to the phase shift registered by the detector is of second order when one uses amplitude modulation technique. Therefore as far as $\delta A/A\gg (\delta \theta)^2$, the phase shift effect is negligible.

Even without hypothetical ALPs, the vacuum is itself birefringent when presence of a magnetic field due to the Heisenberg-Euler term $\frac{\alpha^2}{90m_e^4}[(F_{\mu\nu}F^{\mu\nu})^2 +\frac{7}{4}(F_{\mu\nu}\tilde{F}^{\mu\nu})^2]$). For $B\sim10$T, $L\sim 10$m, $\omega\sim {\rm eV}$, this QED effect of the phase shift is of ${\cal O}(10^{-14})$ so it is registered by the detector of order ${\cal O}(10^{-28})$ which is negligible.

In our proposal, an amplitude modulated laser beam is divided by a beamsplitter into two beams with equal intensity. One beam passes through a region permeated by a constant magnetic field. For the detection of pseudoscalar ALPs, we will set the beam linearly polarized in the direction of the magnetic field (For the detection of scalar ALPs, the polarization should be perpendicular to the magnetic field).  The two beams are then recombined at the detector, and in the presence of a conversion due to the hypothetical particles, the amplitude reduction leads to interference, which can be detected.

The length of the path traversed by one beam is designed slightly different from the other one, so that at the detector the two beams would be out of phase by $\pi$ if the magnetic field has been absent. The purpose for this arrangement is to reduce the background.

Let the path lengths of the two arms be $L_x$ and $L_y$, and the state of the laser after amplitude modulation is
\begin{equation}
\vec E_{in} = \vec E_{0} (1+\beta \sin \omega_m t) e^{i\omega t},
\end{equation}
where $\beta$ is a constant, $\vec E_0$ is the initial electric field at $t=0$, and $\omega$ is the frequency of the laser.  The amplitude is modulated at a frequency $\omega_m$.  This can be recast as
\begin{equation} \label{Einitial}
\vec E_{in} = \vec E_{0} \left(e^{i\omega t} +\frac{\beta}{2i}e^{i(\omega+\omega_m)t} - \frac{\beta}{2i}e^{i(\omega-\omega_m)t}\right),
\end{equation}
where the first term is referred to as the ``carrier'', and the latter two as ``sidebands''.

The state of the carrier after recombination at the detector is given by
\begin{eqnarray} \nonumber
\vec E_{carrier} &=& -\frac{\vec{E}_0}{2} e^{i(\omega t + 2 kL)} \\
&& \times \bigg[2i \sin k\Delta L - (\frac{\delta A}{A}+i\delta \theta ) e^{-i k\Delta L}\bigg],
\end{eqnarray}
where $k = \omega /c$ is the wavenumber of the laser photons, $A = |\vec E_0|$, $\Delta L = L_x-L_y$ is the length difference between the two arms, and $L = (L_x + L_y)/2$ is the average.  As mentioned, we will choose $k\Delta L=\pi$, so that the detector operates at a dark fringe, in order to eliminate the background.  This leads to
\begin{equation}
\vec E_{carrier} = {e^{i(\omega t + 2kL)}\over 2} (\frac{\delta A}{A}+i\delta \theta ) \vec{E}_0.
\end{equation}
Note that without the aid of the sidebands, this would be the entire signal.  While the background is eliminated, the intensity ($\sim \vec E^2$) is of $\mathcal{O}(g_{a\gamma\gamma}^4)$.  This loss in sensitivity, as we will see, can be recovered by using the sidebands.

Meanwhile, the sidebands (second and third terms of \eqref{Einitial}) are described by
\begin{eqnarray} \nonumber
\vec E_{\pm} &=& \vec E_0\beta e^{i(\omega t + 2kL)}e^{\pm i(\omega_m t + 2\omega_m L/c)} \\
&& \times \bigg[\sin\frac{\omega_m \Delta L}{c} \mp i(\frac{\delta A}{A}+i\delta \theta) {e^{\mp i \omega_m \Delta L/c}\over 2}\bigg],
\end{eqnarray}
where the subscripts $+$ and $-$ denote respectively the sideband components of frequency $\omega+\omega_m$ and $\omega-\omega_m$.

If we set $\omega_m  \approx \pi c/2\Delta L$, the total electric field at the detector is obtained by adding that of the carrier and sidebands:
\begin{eqnarray} \nonumber
\vec E &=& \vec E_0 e^{i(\omega t+2kL)}\bigg({1\over 2}(\frac{\delta A}{A}+i\delta \theta) \\
&& + \beta \left(2-(\frac{\delta A}{A}+i\delta \theta)\right)\cos\left[\omega_m t + \frac{2\omega_m L}{c}\right]\bigg)
\end{eqnarray}
Note that this particular value of $\omega_m$ is chosen to maximize the signal.  Since $\omega_m \rightarrow n\omega_m$ and $k\Delta L \rightarrow n\pi$ (for $n$ an odd integer) are equally valid choices, the experimenter has much freedom in choosing a suitable value for $\omega_m$ that is experimentally feasible.

Hence, the power $P$ that falls on the detector is
\begin{eqnarray} \nonumber
P &=& P_{in}\bigg\{ \frac{(\delta A/A)^2+\delta \theta ^2}{4}
+ \frac{\beta^2(4-4{\delta A\over A}+{\delta A^2\over A^2}+\delta\theta ^2)}{2} \\ \nonumber
&+& \beta (2{\delta A\over A}- {\delta A^2\over A^2}+{\delta\theta^2\over 2})\cos \left[\omega_m \left(t + \frac{2 L}{c}\right)\right] \\
&+& \frac{\beta^2(4-4 \frac{\delta A}{A}+ \frac{\delta A^2}{A^2}+\delta \theta ^2)}{2}\cos \left[2\omega_m \left( t + \frac{2L}{c}\right)\right]\bigg\}\nonumber\\.
\end{eqnarray}

Thus the power has a dc component (first line), and two ac components with frequencies $\omega_m$ and $2\omega_m$.  If we multiply this with the oscillator voltage that drives the Pockels cell (plus an appropriate phase shift) via a mixer, we can extract the component of frequency $\omega_m$.  Neglecting the second-order contributions, the time-averaged output power of the mixer is given by
\begin{eqnarray}
P_{out} &=& \frac{1}{T} \int_T 2 P_{in}\beta \mathcal{G} \left(\frac{\delta A}{A}\right) \cos^2 \left(\omega_m t\right) \\
&=& \frac{P_{in} \beta\mathcal{G}\delta A}{A}
\end{eqnarray}
where $\mathcal{G}$ is the gain of the detector and $T$ is taken to be sufficiently long to ensure that the time-averaging is accurate.  Hence, the output signal is proportional to $g_{a\gamma\gamma}^2$ for axions or ALPs.

In this analysis we choose to modulate the amplitude, rather than the phase, of the photons so the result will not be spoiled by the QED effect. In principle, we could instead modulate the phase, in which case the change in intensity registered by the detector would be primarily a consequence of the phase shift instead of the amplitude reduction.  The corresponding analysis is highly analogous and will not be repeated here. For an experiment mainly interested in measuring the QED effect, the phase modulation should be employed.

Despite the improvement in signal size, the use of interferometers is inevitably accompanied by the presence of shot noise. This limits the resolution of the interferometer therefore reducing the sensitivity to $g_{a\gamma\gamma}$ in our set up.

For a laser beam consisting of $N$ incoming photons, we expect the shot noise has a magnitude of $\sqrt{N}$.  The signal-to-noise ratio is thus reduced to $(g_{a\gamma\gamma}BL)^2 N/\sqrt{N}$.  In the case of a non-detection, this allows us to constrain the axion-photon coupling to $g_{a\gamma\gamma, max}< (BL)^{-1}N^{-1/4}$, which is what can be achieved by conventional photon-regeneration experiments. (In photon-regeneration experiments, the signal is much smaller, of $\mathcal{O}(g_{a\gamma\gamma}^4 N)$, so dark count rate can be a problem.)

Our setup admits a straightforward implementation of squeezed light, which can help reduce shot noise. Shot noise can be viewed as the beating of the input laser with the vacuum fluctuations entering the other side of the beam splitter. The conception of reducing shot noise by injecting squeezed light was first suggested by~\cite{caves}. Implementations of squeezed light together with using power recycling and sidebands have been demonstrated and a $10$dB shot noise reduction was achieved~\cite{squeezwithcave2}. A $10{\rm dB}$ suppression of shot noise can result in a $10^{1/2}$ improvement of the constraint to $g_{a\gamma\gamma}$.

To further boost the sensitivity, we can incorporate in our setup optical delay lines or Fabry-Perot cavities to enhance the signal by a factor of $n$, where $n$ is the number of times the laser beam is folded. The resultant improvement in our ability to constrain $g_{a\gamma\gamma}$ is of order $n^{1/2}\sim 1000$ v.s. $n^{1/4}\sim 10^{1.5}$ in the photon regeneration experiment. Combined, the use of squeezed light and optical delay lines results in a gain in the sensitivity to $g_{a\gamma\gamma}$ of $10^{2}$ over a power recycled photon regeneration experiment for current techniques.

If we use $n\sim 10^6$, $B\sim 10{\rm T}$, $L\sim 10{\rm m}$ with a $10$W $(\lambda=1\mu{\rm m})$ laser, after $240$ hours running, the experiment can exclude ALPs with $g_{a\gamma\gamma}>0.9\times 10^{-11}{\rm GeV^{-1} }$ and $m_a\ll10^{-4}{\rm eV}$ to $5\sigma$ significance. If one also employs squeezed-light laser which improves signal-to-noise ratio by $10$dB with a similar setup, the exclusion limit can reach $g_{a\gamma\gamma}\sim3\times10^{-12}{\rm GeV^{-1}}$.

While we aim for the detection of ALPs in the discussion, our design is theoretically applicable to any sub-eV particles that couple to two photons.

\begin{footnotesize}

\end{footnotesize}



\begin{thebibliography}{99}
\bibitem{Sikivie:1983ip}
  P.~Sikivie,
  Phys.\ Rev.\ Lett.\  {\bf 51}, 1415 (1983).

\bibitem{Raffelt:1987im}
  G.~Raffelt, L.~Stodolsky,
  Phys.\ Rev.\ D {\bf 37}, 1237 (1988).

\bibitem{tamyang}
H. Tam and Q. Yang, Phys. \ Lett. B {\bf 716} (2012) 435-440.

\bibitem{caves}
C. M. Caves, Phys. \ Rev.\ D {\bf 23}, 1693 (1981)

\bibitem{squeezwithcave2}
H. Vahlbruch, {\cal et al.} Phys. \ Rev.\ Lett. {\bf 100}, 033602 (2008)
\end{thebibliography}
\end{document}